\documentclass[final,1p,times,twocolumn]{elsarticle}
\usepackage{amssymb}
\usepackage[T1]{fontenc}
\usepackage[utf8]{inputenc}
\usepackage{amsfonts,amsbsy,amssymb,amsmath,graphicx,float} 
\usepackage{mathtools}
\usepackage{color}
\usepackage[style=base]{caption}
\graphicspath{figures/}
\usepackage{graphicx}
\usepackage{subfigure}

\journal{Physica D: Nonlinear Phenomena}

\begin{document}

\begin{frontmatter}

\title{Analysis of the spread of SARS-CoV-2 in a hospital isolation room using CFD and Lagrangian Coherent Structures}

\author[1,2]{Narjisse Amahjour\corref{cor1}}
\cortext[cor1]{narjisse.amahjour@gmail.com}
\author[1,3]{Guillermo Garc{\'i}a-S{\'a}nchez}
\author[4]{Makrina Agaoglou}
\author[1]{Ana Maria Mancho}
\affiliation[1]{organization={Instituto de Ciencias Matemáticas, CSIC},
			addressline={C/ Nicolás Cabrera 15, Campus Cantoblanco},
			city={Madrid},
			postcode={28049},
			country=Spain}

\affiliation[2]{organization={Department of Physics, Faculty of Sciences, Abdelmalek Essaadi University},
			city={Tetouan},
			postcode={93002},
			country=Morocco}

\affiliation[3]{organization={Escuela T{\'e}cnica Superior de Ingenieros de Telecomunicación, Universidad Polit{\'e}cnica de Madrid},
			addressline={Av. Complutense, 30},
			city={Madrid},
			postcode={28040},
			country=Spain}
\affiliation[4]{organization={Departamento de Matem{\'a}tica Aplicada a la Ingenier{\'i}a Industrial, Escuela T{\'e}cnica Superior de Ingenieros Industriales, Universidad Polit{\'e}cnica de Madrid, c/ Jos{\'e} Guti{\'e}rrez Abascal},
			addressline={c/ Jos{\'e} Guti{\'e}rrez Abascal, 2},
			city={Madrid},
			postcode={28006},
			country=Spain}

% \affiliation[1]{Instituto de Ciencias Matemáticas, CSIC, C/ Nicolás Cabrera 15, Campus Cantoblanco, 28049, Madrid, Spain.}
% \affiliation[2]{Department of Physics, Faculty of Sciences, Abdelmalek Essaadi University, 93002, Tetouan, Morocco}
% \affiliation[3]{Escuela T{\'e}cnica Superior de Ingenieros de Telecomunicación, Universidad Polit{\'e}cnica de Madrid, Av. Complutense, 30, Madrid,28040,Spain}       
% \affiliation[4]{Departamento de Matem{\'a}tica Aplicada a la Ingenier{\'i}a Industrial, Escuela T{\'e}cnica Superior de Ingenieros Industriales, Universidad Polit{\'e}cnica de Madrid, c/ Jos{\'e} Guti{\'e}rrez Abascal, 2, Madrid, 28006,Spain}

\begin{abstract}
This research paper presents an analysis of the propagation of the SARS-CoV-2, or other similar pathogens, in a hospital isolation room using computational fluid dynamics (CFD) and Lagrangian Coherent Structures (LCS). The study investigates the airflow dispersion and droplets in the room under air conditioning vent and sanitizer conditions. The CFD simulation results show that  both the air conditioner and sanitizer systems significantly influence the dispersion of the virus in the room. The use of LCS  enables acquiring a deep understanding of the dispersion of suspended  particles, providing insights into the mechanisms of virus transmission. The findings of this study could help in developing strategies for improving the design and operation of isolation rooms to minimize the likelihood of virus dissemination within hospitals.
\end{abstract}

% %Research highlights
% \begin{highlights}
% \item Research highlight 1
% \item Research highlight 2
% \end{highlights}

\begin{keyword}
CFD \sep Covid-19 \sep hospital isolation room \sep air-conditioner \sep sanitizer \sep Lagrangian Coherent Structures
\end{keyword}

\end{frontmatter}

\section{Introduction}
The onset of the COVID-19 pandemic, which is induced by the SARS-CoV-2 virus, was first identified in Wuhan, China, in late 2019, resulting in a widespread outbreak of respiratory illness with severe implications. The virus can cause severe respiratory distress and death if not treated appropriately \cite{Setti2020}. The pandemic has not only resulted in significant loss of life but has also had a profound impact on the economies of affected nations.
Transmission of the virus occurs through respiratory droplets and aerosols produced by an infected individual during activities such as breathing, coughing, sneezing, singing, shouting, or speaking. Droplets emitted from infected individuals come in varying sizes, with larger droplets falling quickly to the ground, while smaller droplets, referred to as aerosols, can remain suspended in the air, particularly in indoor environments. 
The extent to which droplets of different sizes cause infection is not fully comprehended. These droplets or aerosols may enter another individual's nose, mouth, or eyes directly, or be inhaled into their airways and lungs. It has been observed that respiratory droplets are typically larger than 5-10 micrometers in diameter. As they evaporate, some of them generate microscopic aerosols, with a diameter of fewer than 5 micrometers, which remain in suspension indoors. These aerosols can be produced by normal breathing and conversation, posing a risk of contamination by inhalation if they contain the virus in sufficient quantity.

It is crucial to note that the smaller droplets, or aerosols, are considered to be more dangerous as they can be advected by the air and remain suspended for longer periods of time, thereby increasing the potential for transmission. Therefore, one of the most important steps in preventing and treating infectious diseases caused by SARS-CoV-2, or other pathogens propagating similarly, is to research the spread of contamination in indoor environments, particularly in intensive care units, in order to identify and implement effective mitigation strategies.

Various scientific studies have utilized CFD modeling to examine the behavior of airflow in isolation rooms and to assess the efficacy of different interventions in various indoor settings, including ventilation systems, surgical masks, and sanitizers. 

Verma et al. (2018) \cite{Verma2018} used CFD techniques to study the evolution of particle dispersion in a single-bed hospital room. The study aimed to understand the transmission routes of the illness and the impact of air change rate (ACR) on airborne disease clearance. The findings showed that higher ACRs result in faster clearance of airborne diseases and the distance to the outlet is a crucial factor in infection management. Bhatia et al.(2020) \cite{Bhatia2020} used CFD simulations to visualize droplets exhaled by an infected individual in a 10m² cabin space. The research discovered that around 75\% of the droplets emitted by an infected individual disperse inside the aircraft cabin, travelling up to two meters behind the individual and mixing with the airflow within 20 seconds. As a result, passengers seated behind the infected individual are at risk of infection.
Using CFD, Bhattacharya et al.(2020) \cite{Bhattacharya2020} conducted research to evaluate the effectiveness of aerosol sanitizer and conditioned air released from air conditioners in eliminating the SARS-CoV-2 virus in isolation rooms. The study showed that high turbulence fields within the room could aid in the dispersion of sanitizer, leading to the eradication of the virus.

Leonard et al.(2020)\cite{Leonard2020}  employed CFD to investigate the effectiveness of a simple surgical mask in reducing the aerosol spread of SARS-CoV-2 during high-velocity nasal insufflation. The study revealed that the surgical mask filtered and collected 88.8\% of the total respiratory particles generated by nasal insufflation, with about 96.5\% of the weighted particles filtered. Furthermore, the majority of the escaped particles during the high-velocity nasal insufflation were settled within 1 meter of the patient's face, and only 2.97\% of them were able to travel a significant distance.
Arjmandi et al. (2022) \cite{Arjmandi2022} used CFD to examine how various factors affect the effectiveness of ventilation systems in reducing the transmission of SARS-CoV-2 in an intensive care unit (ICU) room. The study identified two ventilation criteria, namely the dimensionless timescale (T) and extraction timescale ($\tau$), and concluded that poor ventilation systems can increase the risk of contamination, as they behave like a perfectly stirred reactor.

Generally, these studies used CFD approach to understand the behavior of respiratory droplets, the impact of different parameters such as air change rate and distance to the outlet, and the efficacy of different interventions such as ventilation systems, surgical masks, and sanitizers in minimizing the spread of SARS-CoV-2 in indoor environments.

This work aims to investigate the behavior of flow transport inside a 3D isolation hospital room, with a particular focus on the area around the patient bed. The study specifically investigates the impact of both the air conditioner and sanitizer on the transport of suspended particles in the room. Particularly, the focus is on examining whether the circulation established between these ventilation and disinfecting mechanisms is effective in evacuating and/or purifying the air around the head of a patient infected with SARS-CoV-2. To achieve these objectives, we  combine the use of  a numerical simulation of the airflow   and analysis techniques based on Lagrangian Coherent Structures (LCS).  
The CFD simulation,  under prescribed inflow-outflow conditions imposed by the air conditioner and sanitizer, generated turbulent velocity fields.  The numerical simulation of turbulent flows is indeed challenging due to the intricate nature of motion occurring across multiple scales. To tackle this challenge, we utilize the Reynolds-averaged Navier–Stokes equations (RANS equations). 
The turbulent flow patterns sustain the transport  of suspended particles in the room. 
The study of transport processes in three-dimensional (3D) flows poses difficulties as fluid parcels within them can follow highly intricate trajectories, even in well-controlled experiments \cite{speetjens, wiggins2010}. LCS have proven to be a successful tool for investigating 3D transport in various flow scenarios \cite{niang2020, renzo2023}.
They extract ordered transport  patterns  from which we identified  regions of the flow with different sources or fates. 
This approach allows for gaining insights into the pathways of particle transport and virus dispersion in the room.

The structure of the paper is as follows. Section 2 discusses the methodology used to describe the turbulent flow in the isolation room. Section 3 analyzes transport in the data set. Finally, Section 4 summarizes the conclusions and perspectives.

\begin{figure}[!htbp]
\centering
\includegraphics[scale=0.58]{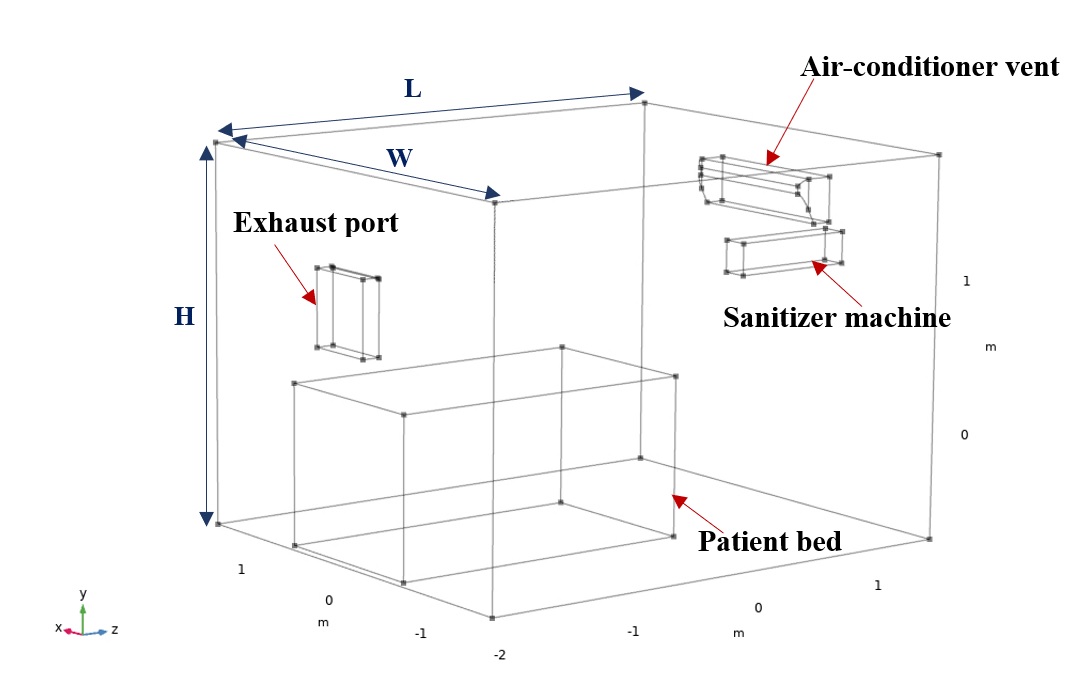}
\caption{ The geometry and domain of the hospital isolation room for computational analysis.} 
\label{room}
\end{figure}

\section{ A 3D turbulent flow model of the isolation hospital room} \label{data}
%A 3-dimensional turbulent flow was  simulated in an isolation hospital room   using the COMSOL software. This software has been extensively used for modeling and solving different scientific and engineering applications \cite{Manjesh2022, Sofi2023, Cheng2022, narjisse2017analysis, narjisse2020computational, narjisse2021assessment, COMSOL}.

%COMSOL Multiphysics is an excellent choice for CFD simulation due to its advanced capabilities in modeling complex fluid dynamics. CFD simulation involves the use of the Finite Element method to analyze the behavior of fluids in a variety of applications, such as aerodynamics, chemical engineering, and biomedical research. COMSOL Multiphysics provides a powerful platform for these simulations, allowing users to model and analyze a wide range of physical phenomena, including turbulent flows, multiphase flows, and heat transfer. The software's intuitive interface and extensive library of pre-built models and simulations make it easy to use for both novice and advanced users. In addition, COMSOL Multiphysics allows for customization and modification of simulations, giving researchers the flexibility to tailor their simulations to their specific needs. With its advanced capabilities and user-friendly interface, COMSOL Multiphysics is an ideal choice for CFD simulation in a wide range of industries and applications.

A 3-dimensional turbulent flow was  simulated in an isolation hospital room. The  domain of the isolation hospital room is shown in Figure \ref{room}. This is a generic geometry similar to those used in other studies exploring general hospital room settings \cite{Arjmandi2022}, but with certain distinctions. Specifically, we adopted the layout of a standard isolation room designed for a single patient. The setup includes an air-conditioner, an exhaust port, a patient bed, and a sanitizer machine, the latter of which was not considered in the references \cite{Arjmandi2022, sahul2019}. Table \ref{dimension} summarizes the dimensions used in the structural modeling process.
\begin{table}[!htbp]
\centering
\begin{tabular}{c|c|c|c}
Area name &  Length(m)  &  Width(m) & Height(m) \\
\hline
Isolation room & 3.5 & 3.0 & 2.5 \\
Air-conditioner vent & 1.1 & 0.18 & 0.3 \\
Exhaust port & 0.49 & 0.5 & 0.12 \\
Patient bed & 2.13 & 1.2 & 1.05\\
Sanitizer machine & 0.8 & 0.16 & 0.2 \\
\hline
\end{tabular}
\caption{3D model parameters}
\label{dimension}
\end{table}

\subsection{The governing equations}
The equations that govern the flow within the room  are the continuity and momentum equations, which are based on the principle of conservation of mass and conservation of momentum, respectively. We do not solve the full 3D equations in a turbulent regime; instead, we utilize the Reynolds-Averaged Navier-Stokes (RANS) equations. This approach effectively separates the fluid motion into a mean  and a fluctuating component. The mean component represents the overall behavior of the fluid, while the fluctuating component represents the turbulent fluctuations.

\textbf{Continuity and Momentum equation:} 
The RANS equations for a steady state incompressible flow give the following descriptions of the conservation of mass and momentum:
\begin{equation}
\frac{\partial \overline{u_i}}{\partial x_i} =0 
\end{equation}
\begin{equation}
\rho \frac{\partial\overline{u_i}}{\partial x_i} +\rho\overline{u_j} \frac{\partial \overline {u_i}}{\partial x_j} =- \frac{\partial \overline{p}}{\partial x_i}+ \frac{\partial}{\partial x_j} (\mu(\frac{\partial\overline{u_i}}{\partial x_j}+\frac{\partial\overline{u_j}}{\partial x_j})-\overline{\rho u'_i u'_j})+\overline{f} 
\end{equation}
where $\overline{u}$  and  $\overline{p}$ are the mean flow velocity and pressure, $\rho$ is the fluid density,   $\overline{f}$ is to define additional forces, like gravity, and $ \mu$ is the dynamic viscosity. Here $i$ is the subindex that runs across the spatial coordinates taking  values 1, 2, and 3. The Reynolds stress term $-\overline{\rho u'_i u'_j}$, which represents the fluctuating part of the momentum transport in turbulent flows, can be determined using the Boussinesq approximation presented in equation (3):
\begin{equation}
\tau_{ij}=-\overline{\rho u'_i u'_j}= \mu_t(\frac{\partial\overline{u_i}}{\partial x_j}+\frac{\partial\overline{u_j}}{\partial x_i})-\frac{2}{3}\rho k \delta_{ij}
\end{equation}
where $\delta_{ij}$ is the Kronecker function and $\mu_t$ is the turbulent viscosity.

\textbf{Turbulence model}: We considered the  $k-\epsilon$ turbulence model implemented in COMSOL, which is one of the most commonly used turbulence models.  
It consists of two transport equations for turbulent kinetic $k$ and dissipation energy $\epsilon $   written as:

\begin{equation}
\rho\overline{u_i}\frac{\partial k}{\partial x_j}= \frac{\partial}{\partial x_i} [(\mu+ \frac{\mu_t}{\sigma_k})\frac{\partial k}{\partial x_i}] +P_k - \rho \epsilon
\end{equation}
\begin{equation}
\rho\overline{u_i}\frac{\partial \epsilon}{\partial x_j} 
= \frac{\partial}{\partial x_i}[(\mu+ \frac{\mu_t}
{\sigma_\epsilon}) \frac{\partial \epsilon}{\partial 
x_i}]+ \sqrt{2}C_{\epsilon_1}S_{ij}\epsilon-
C_{\epsilon2}\rho\frac{\epsilon^{2}}{k+\sqrt{\nu\epsilon}}
\end{equation}
where the turbulent viscosity is defined by:
\begin{equation}
\mu_t=\rho C_\mu \frac{k^{2}}{\epsilon}
\end{equation}
with $C_\mu$ is determined by:
\begin{equation}
C_\mu=\frac{1}{A_0+A_s\frac{kU^*}{\epsilon}}
\end{equation}
\begin{equation}
U^*=\sqrt{S_{ij} S_{ij} + \Omega_{ij} \Omega_{ij}}
\end{equation}
\begin{equation}
\Omega_{ij}= \overline{\Omega_{ij}}-\epsilon_{ijk} \omega_k -2\epsilon_{ijk} \omega_k
\end{equation}

where $\overline{\Omega_{ij}} $  is the average rate of rotation tensor and $ \omega_k $ is the angular velocity.

The two constants $A_0 $ and $ A_s $ are calculated as follows:
\begin{equation}
A_0 =4,
A_s=\sqrt6 \cos{\phi}
\end{equation}
\begin{equation}
\phi=\frac{1}{3}\arccos(\min(\max(\sqrt{6}W,1),1))
\end{equation}
\begin{equation}
W=\frac{S_{ij} S_{jk} S_{ki}}{S^{2}}
\end{equation}
$C_{\epsilon 1} $  is determined as:
\begin{equation}
C_{\epsilon 1}=\max(\frac{\eta}{5+\eta}, 0.43)
\end{equation}
where:
\begin{equation}
\eta=S(\frac{k}{\epsilon})
\end{equation}

The closure constants $C_2 , \sigma_k , \sigma_\epsilon $ were defined by \cite{Shih1995} as below:

$C_2=1.9, \sigma_k=1.0 , \sigma_\epsilon=1.2$

\subsection{The boundary conditions}
The boundary conditions used to solve the above equations are summarized in Table \ref{BC}.
The computational domain contained air as a fluid, while the solid domains were designated as no-slip walls. 
To set the inlet boundary conditions, a velocity inlet, turbulent kinetic energy $k_{in}$, and turbulent dissipation energy $\epsilon_{in}$ were specified. The properties of air were obtained from the COMSOL library. These turbulence parameters are implemented in COMSOL using the following  expressions:

\begin{equation}
k_{in}=\frac{3}{2}(U_{in} I_T)^{2}
\end{equation}

\begin{equation}
\epsilon_{in}=\frac{C_\mu^\frac{3}{4}k^\frac{3}{2}}{L_T}
\end{equation}

\begin{table}[!htbp]
\resizebox{\columnwidth}{!}{
\begin{tabular}{|l|l|l|}
\hline
Description & Type     & Flow                    \\ \hline
Inlet &
Velocity and turbulence parameters, $k_{i_n}$, $\epsilon_{i_n}$ &
\begin{tabular}[c]{@{}l@{}}- Air-conditioner : u=4m/s , turbulent length scale = 0.01 , turbulent intensity = 0.05.\\ - Sanitizer machine : u=1.5 m/s turbulent length scale $(L_T)$=0.01,  turbulent intensity $(I_T)$=0.05\end{tabular} \\ \hline
Outlet      & Pressure & Exhaust port : $p_0$    \\ \hline
Walls       & No-slip  & All walls , patient bed \\ \hline
\end{tabular}%
}
\caption{Summary of boundary conditions}
\label{BC}
\end{table}

%simulate the airflow physics.
\subsubsection{The numerical implementation} %\label{data}

\begin{figure}[!htbp]
 \centering
\includegraphics[scale=0.8]{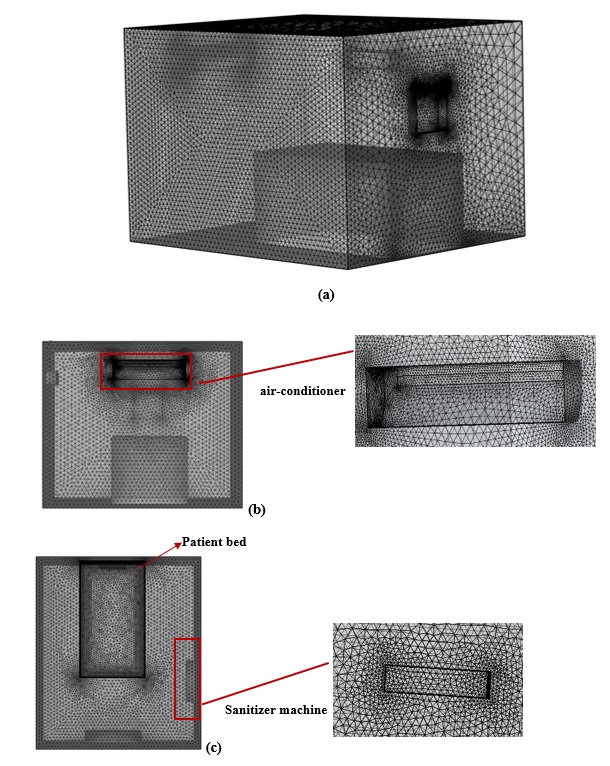}
\caption{Details of numerical CFD simulation grid a) grid of isolation room b)  grid of $x$-$y$ view and c) grid of $x$-$z$ view  } 
\label{mesh room}
\end{figure}

Numerical computations of the turbulent flow were conducted using COMSOL Multiphysics software \cite{COMSOL}. COMSOL is a versatile software widely employed in the scientific literature for the numerical modeling of phenomena such as plasmas \cite{brez2015}, heat transfer \cite{z2012}, shallow streams \cite{n2017}, turbulence \cite{m2012,Manjesh2022,Sofi2023,Cheng2022,narjisse2021assessment}, and more.
COMSOL     
uses a Finite Element Method (FEM) to numerically  create the mesh and to solve the  proposed equations. In this work, the simulations using COMSOL Multiphysics 5.4, have been  run on an Intel®Core (TM) i7-6700HQ CPU @ 2.60GHz with available memory of 12 GB and a 64-bit system. COMSOL estimates the size of the numerical problem by reporting the number of degrees of freedom. The mesh employed in our study consisted of tetrahedral shapes in 3D configuration. The specific mesh sizes varied throughout the domain. A refined mesh was implemented in critical areas such as the inlet, outlet, and near the patient's bed to enhance result precision. The mesh consisted of 1752243 elements, including 70390 boundary elements and 2554 edge elements. Figure \ref{mesh room} provides a detailed illustration of the mesh on the main isolation room, where a combination of hexahedral and tetrahedral elements was utilized, specifically, the patient bed, HVAC system, exhaust port, and sanitizer machine were represented using hexahedral elements to accommodate their complex geometries, and the remaining regions of the computational volumes were meshed using tetrahedral elements, in addition, it shows a magnification of the key components, air conditioning, and sanitizer.
In this work, a user-controlled mesh based on the fluid dynamics option was used, which is built from finite elements of different types and sizes. COMSOL offers nine different mesh size types that vary from extremely fine to extremely coarse \cite{COMSOL}. In this study, we assessed our current model by employing four different mesh sizes, from an extra coarse mesh to a finer mesh. As the mesh resolution increases, the quality of the solutions improves; however, this improvement comes at the cost of longer run times and increased RAM usage. For this reason, in this study, we based on the mesh independence test by evaluating the variation of the velocity parameter through the domain for the four different mesh sizes mentioned above. This process leads us to ensure that the choice of mesh size does not significantly impact the results. From this point, we choose the normal type mesh which contains 1752243 elements to follow our computation. Mesh quality can be assessed through several measures, including skewness, element aspect ratio…etc. The turbulence variables were separately resolved from momentum and pressure by utilizing a segregated solver. In order to achieve an accurate solution, convergence criteria were set for the residuals: $10^{-5}$ for velocity and pressure, and  $10^{-6}$ for turbulence parameters. These criteria were selected to ensure the achievement of proper results.

\subsection{Results on  flow in the hospital room.}
The results of the CFD simulation of the turbulent airflow inside a hospital isolation room with an air-conditioner, sanitizer, patient bed, and exhaust port are presented next. 
Under the RANS approach, the focus is on the averaged velocity components $\overline{u_i}$, which represent stationary fields. In the RANS approach with the $k-\epsilon$ closure model, the turbulent velocity fluctuations ${u_i}'$ are not explicitly computed. However, the model provides estimates of important quantities such as the turbulent kinetic energy $k$ among others.

\begin{figure*}[!htbp]
 \centering
\includegraphics[scale=0.5]{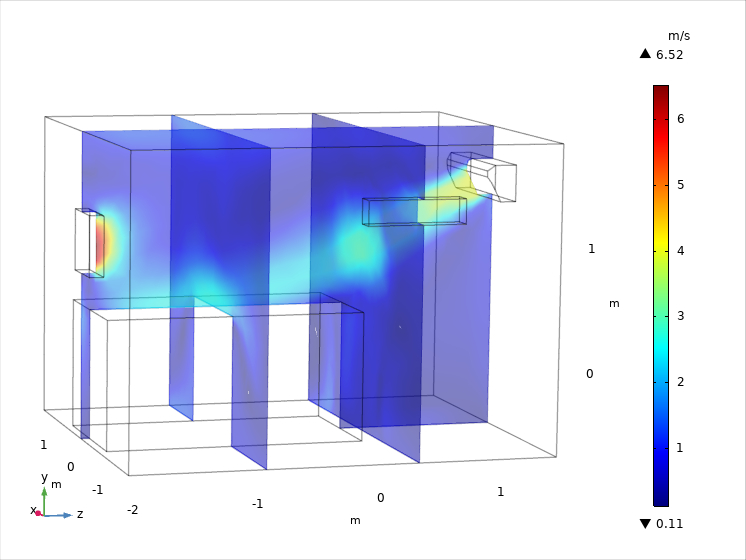}
\caption{ A representation of  $||\mathbf{u}||$ inside the isolation room at cut-planes: $x=0$, $z=-0.9$  and $z=0.35$. }
\label{velocity}
\end{figure*}

In order to make the notation more intuitive, in the rest of the manuscript we will use the notation: $$x_1=x, \,\,\,\,\,\,x_2=y,\,\,\,\,\,\,x_3=z, \,\,\,\,\,\, {\bf x}=(x,y,z),  \,\,\,\,\,\,{\rm and}  \,\,\,\,\,\,\overline{u_1}=u_x, \,\,\,\,\,\,\overline{u_2}=u_y, \,\,\,\,\,\, \overline{u_3}=u_z, \,\,\,\,\,\, {\bf u}=(u_x,u_y,u_z) $$
A representation of the modulus of the averaged velocity field, i.e. $||{\bf u}|| = \sqrt{u_x^2+u_y^2+u_z^2}$, for various section planes inside the isolation room is visible in   Figure \ref{velocity}. These planes are carefully selected to provide a complete 3D overview of the flow in the room. They are placed at $x=0$, which is  a plane that longitudinally intersects the center of the bed and that crosses both the exhaust port and the air conditioning;  a plane at $ z=-0.9$ that is transversal to the bed, and finally a plane at $z=0.35$ that intersects the sanitizer. The figure provides a comprehensive understanding of the complex flow patterns that support the transport behavior of airflow among the essential room elements. The velocity field expression shows the magnitude of the flow velocity, which is the essential field controlling the dispersion of virus particles.
The results of the simulation presented in Figure \ref{velocity} show that at the cut plane $x=0$,  
a high-velocity jet of air is moving towards the patient bed from the air conditioner. The velocity field decreases as it reaches the bed and then flows toward the exhaust port. Additionally, from Figure \ref{velocity}  it can be observed for planes $x=0$ and $z=-0.9$ that the velocity field increase, particularly in the area around the patient bed. This could suggest a higher potential for the spread of SARS-CoV-2 in the room. Indeed, the high velocity may indicate that the fluid particles, including virus particles, are moving more quickly, which may increase their potential to travel longer distances and reach further into the room. This result indicates that the positioning and direction of the air conditioner play a crucial role in the transport behavior of fluid particles in the room. This general overview, roughly deduced from the velocity field, is completed in the next section with a detailed analysis of the transport associated with this velocity field.

\begin{figure*}[!htbp]
 \centering
\includegraphics[scale=0.5]{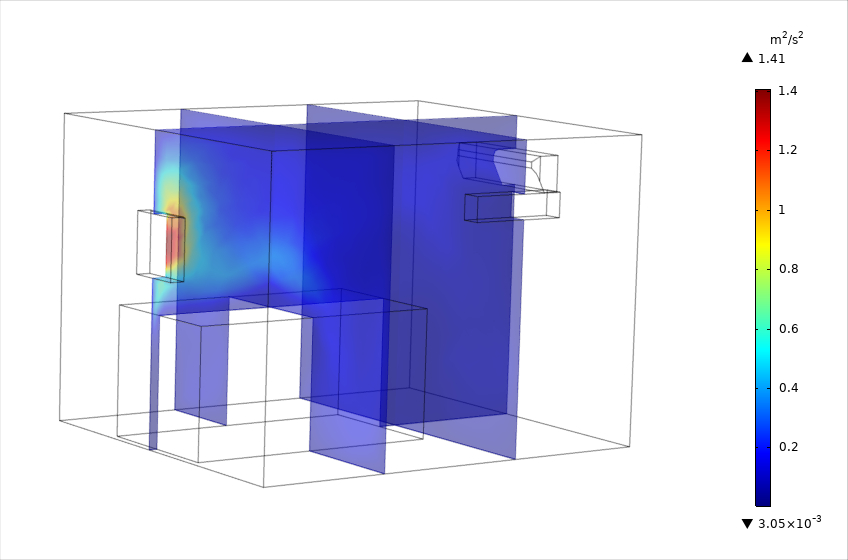}
\caption{A representation of turbulent kinetic energy inside the isolation room at cut-planes: $x=0$, $z=-0.9$  and $z=0.35$. }
\label{TKE}
\end{figure*}

Figure \ref{TKE} shows the turbulent kinetic energy distribution in the isolation hospital room. The representation reveals that the generation of  turbulent fluctuations in the velocity fields is especially pronounced close to the exhaust port. 
%within the isolation room, especially near the patient bed can serve as an effective means of distributing sanitizer throughout the isolation room. This distribution mechanism contributes to the reduction of the SARS-CoV-2 virus.}

\begin{figure*}[!htbp]
 \centering
\includegraphics[scale=0.5]{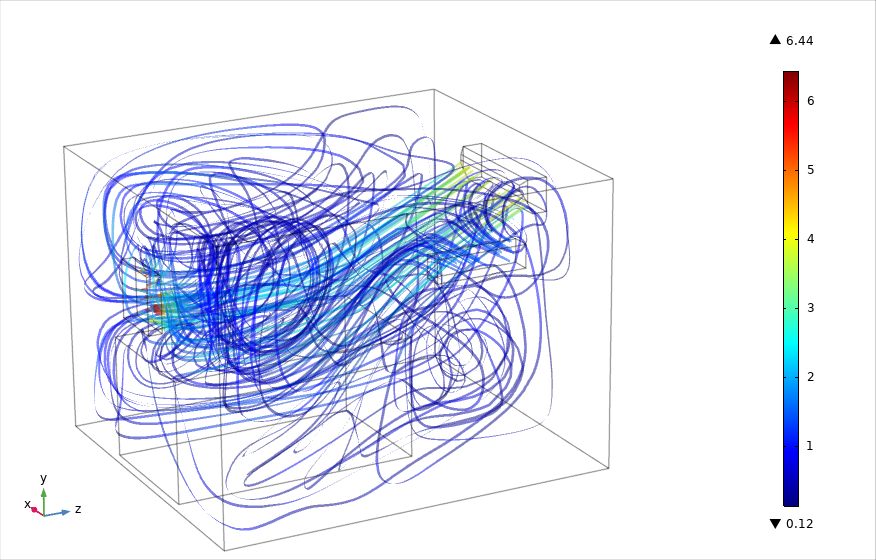}
\caption{A representation of fluid parcels  inside the isolation room. The trajectory paths are colored according to the  value of $||\mathbf{u}||$. }
\label{streamline velocity}
\end{figure*}
\section{Analysis of the Lagrangian Transport}
\label{LDs}
A detailed analysis of the Lagrangian transport induced by ${\bf u}$, requires computing fluid parcel trajectories, which in the advective approach follow the equation:
\begin{equation}
\dfrac{d\mathbf{x}}{dt} = \mathbf{u}(\mathbf{x}) \;,\quad \mathbf{x}=(x,y,z) \;,\; \;
\label{eq:gtp_dynSys}
\end{equation}
Here $\mathbf{u}$  is  the stationary averaged velocity field obtained from the simulations, displayed in Figure \ref{velocity}.  The solutions ${\bf x}(t)$ to the system Eq. \eqref{eq:gtp_dynSys} are the trajectories followed by fluid parcels, and also by virus particles. 
%\nico{Strictly speaking, fluid parcels follow velocities that consist of not only the averaged velocity $\mathbf{u}$ but the total velocity, which also includes the fluctuating component $\mathbf{u}'$ with kinetic energy represented in Figure \ref{TKE}. Given that this component is approximately three orders of magnitude smaller than the averaged component, we do not anticipate it to have a significant impact on the results. We will delve into this discussion later.}

Fluid parcels in 3D flows like this one,
can follow very complicated trajectories.  Even in well-controlled flows such as those in lab experiments \citep{speetjens, wiggins2010} it has been demonstrated that  Lagrangian transport can be very  intricate and chaotic \citep{psc2010,iri}.
This is confirmed in Figure \ref{streamline velocity}, which displays a representation of  a bunch of these trajectories computed by COMSOL. The path of the fluid parcels is colored according to the values of the modulus of the velocity at that point. This  confirms that the velocity of the fluid is highest in the vicinity of the air conditioner and sanitizer, indicating that these systems were effective in generating airflow. However, the velocity decreased as the fluid particles moved further away from the air conditioner and sanitizer, indicating that the transport of fluid particles through the room was not uniform. 
The intricate nature of the results displayed in Figure \ref{streamline velocity} creates a challenge in fully comprehending the transport phenomena. As a result, there is a need to utilize Lagrangian Coherent Structures in order to facilitate a more thorough interpretation, which will be explained in detail below.

\subsection{The perspective of Lagrangian Coherent Structures}
Lagrangian Coherent Structures (LCS) \citep{shadcen2005} provide a geometrical way to look into the trajectories (or solutions) of the dynamical system given by Eq. \eqref{eq:gtp_dynSys}. 
 This perspective is inspired by the work of Poincar\'e, who in the context of celestial mechanics
suggested representing the solutions of systems as Eq.  \eqref{eq:gtp_dynSys} in the phase space and looking for geometric structures that separate regions corresponding to trajectories with qualitatively different dynamic behaviors.
In the case of advection and fluid mechanics, as ours, the phase space of the system  in Eq. \eqref{eq:gtp_dynSys} is the physical space and
LCS act as material barriers that fluid particles cannot cross and  provide a partition of the space separating regions in which fluid parcels behave differently.
%\nico{The results described by Garcia-Garrido et al. in \cite{gg2018} discuss mathematical results regarding  the persistence of these geometric structures in 3D vector fields in the presence of small perturbations. For instance, among these persistent structures there exist normally hyperbolic invariant manifolds (NHIM) and tori.  While describing and identifying such structures in our specific flow is beyond the scope of this paper, it is worth noting that under certain conditions, these geometric structures remain robust against small perturbations. This emphasizes the significance of capturing the major component. Consequently, we anticipate that neglecting the turbulent fluctuating component in the velocity field, given its small magnitude compared to the averaged field, does not significantly distort the structures we describe in the subsequent analysis.}

In this work, LCS are computed by means of the Lagrangian descriptor (LD) known as the $M$ function. This method was introduced in \citep{Madrid2009,mendoza2010,mancho2013lagrangian}, and it already   has been used to visualize  three-dimensional Lagrangian structures  in  3D flows  
\citep{lopesino2017,curbelo2017,curbelo2018b,curbelo2018a,niang2020}.
It also has been implemented in other high dimensional contexts in the field of reaction dynamics \citep{craven2015lagrangian,junginger2016transition,Agaoglou2019}.%,ldbook2020}.  
 The $M$ function is defined as:
 
\begin{equation}
M(\mathbf{x}_{0},t_0,\tau) = \int^{t_0+\tau}_{t_0-\tau}  ||\mathbf{u}(\mathbf{x}(t))|| \; dt
\label{eq:Mp_function}
\end{equation}

 By integrating the trajectory $\mathbf{x}(t)$ both forward and backward in time for a duration of $\tau > 0$, this integral calculates the arclength of the trajectory. Specifically, it measures the distance traveled along the path traced out by $\mathbf{x}(t)$ from an initial point $\mathbf{x}_0=\mathbf{x}(t_0)$ at time $t_0$.
This integral can be split into two terms:
\begin{equation}
M(\mathbf{x}_{0},t_0,\tau) = \int_{t_0-\tau}^{t_0} \mathbf{u}(\mathbf{x}(t)) \; dt   \;+ \int_{t_0}^{t_0+\tau}  \mathbf{u}(\mathbf{x}(t)) \; dt =    M^{(b)}(\mathbf{x}_{0},t_0,\tau)+ M^{(f)}(\mathbf{x}_{0},t_0,\tau)
\label{eq:split}
\end{equation}

For a sufficiently large integration period  $\tau$  both $M^{(b)}$ and $M^{(f)}$ (and therefore also $M$), have a  structure in which  singular features emerge. These patterns mark boundaries between regions in which particles have different qualitative behaviors. This partition is the one of which we will take advantage to further investigate the transport properties of the velocity field in Figure \ref{velocity}. Although we obtain a partition where particles behave differently, a priori it is not possible to know what type of transport is associated with each feature and this is discovered with additional integrations per domain. 
In this article,   we have used $\tau=10$ seconds.  This period of 10 seconds is determined to be sufficient and consistent with the time it takes for fluid parcels in isolation rooms to travel from the air conditioner to the outlet, which is approximately 5 seconds. Shorter integration periods reflect transport associated with fluid particle trajectories that are too short to be of interest, while very large integration periods accumulate too much information on the transport associated with fluid particle trajectories, making it difficult to interpret.

Computing the $M$ function and its interpretation requires integrating trajectories from \eqref{eq:gtp_dynSys}. The algorithm interpolates the 3D scattered velocity field at the trajectory locations using linear interpolation in Cartesian coordinates. Then, the trajectories are integrated forward and backward using a 5th-order Runge-Kutta scheme with a time step of 0.01 seconds. The arc length is computed by adding up the linear segments connecting successive steps of the Runge-Kutta method.
To prevent particles from escaping the room, a boundary of zero velocities was implemented around the room's outer perimeter, simulating a condition where particles approaching the walls would adhere to them. For the bed, a similar approach was employed, with zero velocities introduced within the bed's boundaries. On the other hand, in the vicinity of the air conditioner and sanitizer, no zero velocities were added, as it can be assumed that particles in close proximity to these devices would be expelled.

\subsection{Results}
\begin{figure}[!htbp]
\centering
\subfigure[]{\label{A}\includegraphics[scale=0.41]{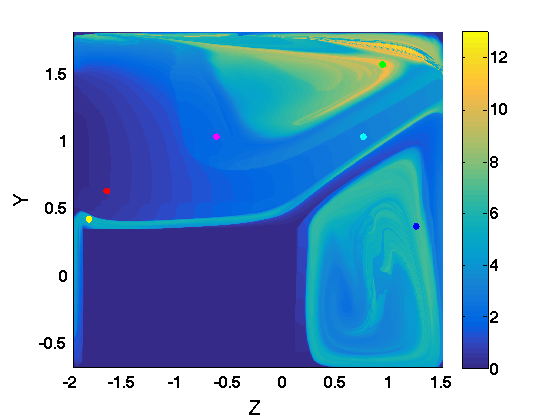}}
\subfigure[]{\label{B}\includegraphics[scale=0.41]{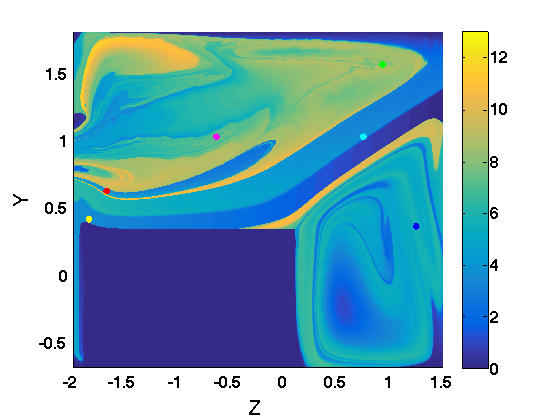}}
\caption{Evaluation of the $M$ function computed in the plane $x$ = 0 for $\tau = 10 $ seconds.  a) forward integration $M^{(f)}$; b) backward integration, $M^{(b)}$. Colored circles are placed in selected domains.}
\label{funM1}
\end{figure}

Figure \ref{funM1} illustrates  the evaluation of $M^{(f)}$ and $M^{(b)}$
on the plane $x=0$ for the Eq. \eqref{eq:gtp_dynSys}.  This representation is obtained for an integration period of $\tau=10$ seconds and, therefore, supports the interpretation of transport within this specific time range. On the other hand, since the vector field in Eq. \eqref{eq:gtp_dynSys} is time-independent or stationary, the geometric structures obtained in this study are also stationary and provide a description of the transport at any given time within a time interval of 10 seconds. 
Colored dots, yellow, red, magenta, green, cyan, and blue are placed in different domains separated by sharp changes in the color code. These sharp changes are material barriers to transport that fluid parcels cannot cross, either in forward or backward time.  In the forward-time representation particles within a domain behave similarly in forward time, while in the backward-time representation,  particles  behave similarly in backward time. The same fluid parcels are displayed both for $M^{(f)}$ and $M^{(b)}$.
These selected fluid parcels  help us to understand how the trajectories evolve around the patient.  We can easily recognize the different motions of a particle and explain what this truly implies in the sense of room ventilation and the spread of the virus.

\begin{figure}[!htbp]
\centering
\includegraphics[scale=0.9]{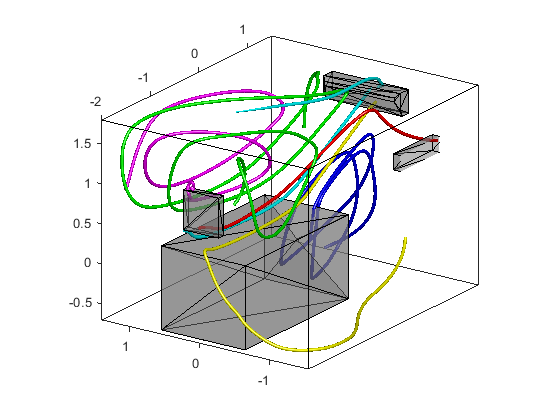}
\caption{ Representation of fluid parcel trajectories associated with the colored circles  placed in selected domains of Figure \ref{funM1}a).}
\label{room2}
\end{figure}

We describe below the different classes of fluid particle behavior associated with each domain. Figure \ref{room2} complements the description of the evolution of the colored trajectories within the room by illustrating their representation over a period of $\tau=10$ seconds both forward and backward in time.
Fluid particles red, magenta, and cyan in Figure \ref{funM1}a) are visible in a domain that corresponds to trajectories that forward time, during this period, leave the room through the exhaust port. This is confirmed  in Figure \ref{room2}. Of these, it is remarkable that the red one starts close to the head of the patient and evolves escaping from the room and similarly will do those that stay in its same domain. The size of this domain offers an indication of the effectiveness of the circulation system in extracting air from the room within a 10-second period.
These colored fluid parcels, however, when looking at panel b) of Figure \ref{funM1} are in different domains, indicating that their behavior differs when traced backward in time. Indeed, going back in time the cyan trajectory flies to the air conditioner and hits it, as Figure  \ref{room2} supports. The red trajectory in the backward-time $M$ structure is in a filamentous elongation of a blue tongue. Its backtracking connects to the sanitizer. In fact,  all the fluid parcels within the tongue-like structure do, giving an idea of how far  volumes of air purified by the sanitizer, are  effectively pushed toward the head of the COVID-19 patient, which is roughly found  in that area. The size of this volume provides an estimate of the effectiveness of the potentially disinfecting particles expelled by the sanitizer in reaching the surrounding area of the patient within the 10-second period.
Finally,  the magenta trajectory, going back in time, remains circulating around the room. This behavior is similar to that of the green trajectory, which is also  in the same backward-time domain as the magenta trajectory. However this green trajectory, in forward time behaves differently from the magenta, as in this time direction the green  also remains circulating around the room (see Figure \ref{room2}) and does not exit through the exhaust port.  
The yellow parcel in forward time is in a domain where parcels surround the bed and stay at the bottom of the room, while in backward time they connect to the air conditioner. Finally, the blue trajectory both in forward and backward time move circling  at the foot of the bed, in the lower part of the room (see also Figure \ref{room2}). All volumes associated with fluid trajectories that do not exit the room in the forward time direction serve as an indicator of the region where ventilation is less effective or ineffective within the 10-second period.

\begin{figure}[!htbp]
\centering
\subfigure[]{\label{A}\includegraphics[scale=0.45]{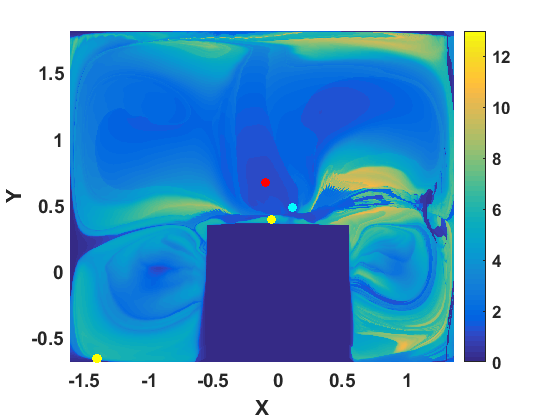}}
\subfigure[]{\label{B}\includegraphics[scale=0.45]{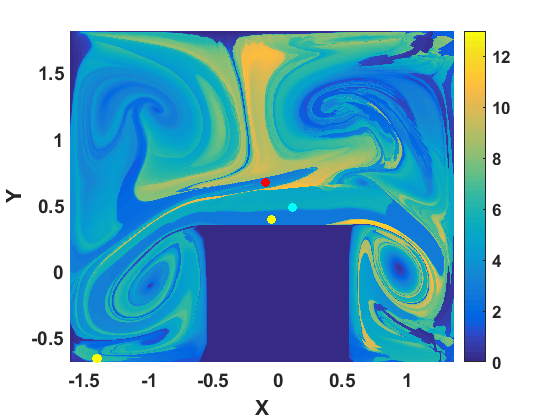}}\\
\subfigure[]{\label{C}\includegraphics[scale=0.45]{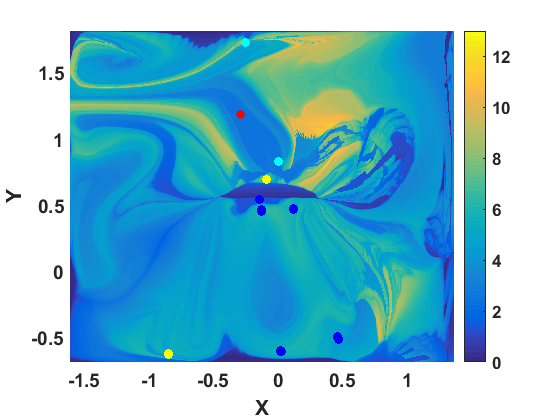}}
\subfigure[]{\label{D}\includegraphics[scale=0.45]{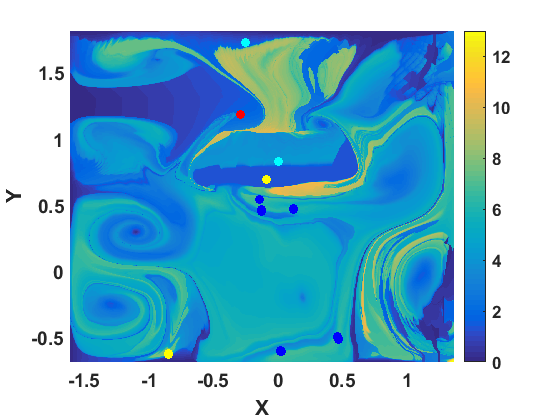}}
\caption{Evaluation of the $M$ function computed  for $\tau = 10 $ seconds in two planes with constant $z$.  a) $M^{(f)}$ at $z=-0.9$; b)  $M^{(b)}$ at $z=-0.9$; c) $M^{(f)}$ at $z=0.35$; d)  $M^{(b)}$ at $z=0.35$. The colored circles are associated with the crossing of the respective trajectories with the planes. They indicate where the regions with different qualitative behaviors are located in the different planes.}
\label{funM2b}
\end{figure}

\begin{figure}[!htbp]
\centering
\subfigure[]{\label{A}\includegraphics[scale=0.35]{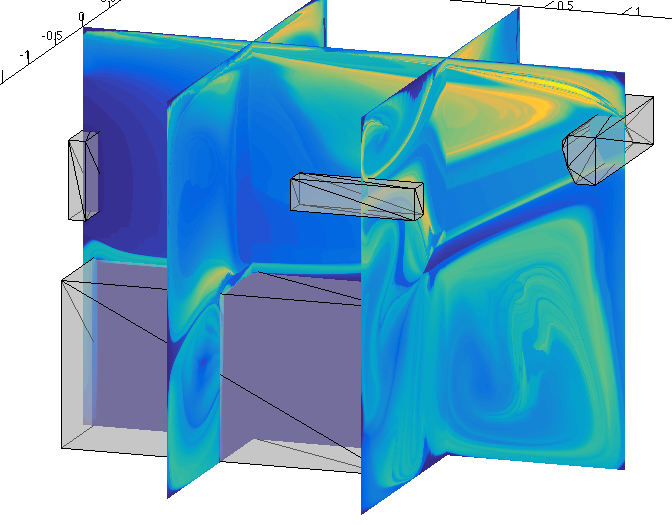}}
\subfigure[]{\label{B}\includegraphics[scale=0.35]{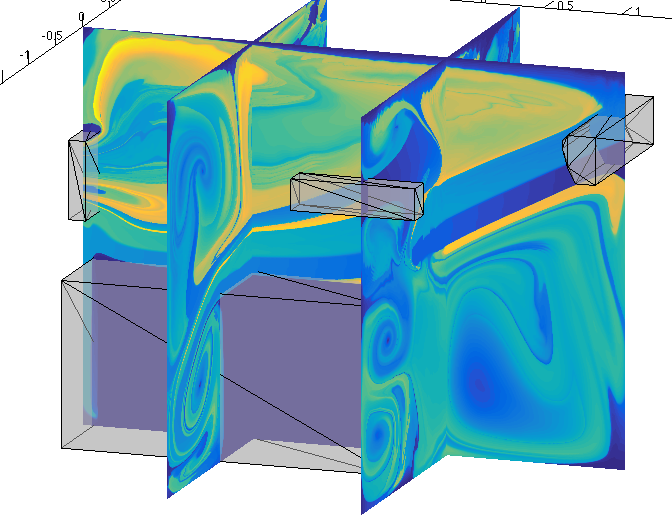}}
 \caption{A 3D representation of  the $M$ function computed  for $\tau = 10 $ at the planes displayed in Figures \ref{funM1} and \ref{funM2b}.  a) $M^{(f)}$; b)  $M^{(b)}$.}
\label{funM3d}
\end{figure}

Figure \ref{funM2b} shows the patterns of $M^{(f)}$ and  $M^{(b)}$ in  the planes $z=-0.9$ and $z=0.35$. These representations support the 3D visualization of the structures identified in Figure \ref{funM1}. The colored circles correspond to where the trajectories associated with each color intersect with the planes. They are therefore linked to the positions in the various planes of the areas with distinct qualitative dynamic behavior. In panels a) and c) the cyan and red dots locate the region in which forward time goes to the exhaust port. The red circle  in panels b) and d), identifies the domain that in backward time goes to the sanitizer. It is remarkable how the LD methodology  is able to capture the distortion of this  volume in different slices of the room.  This ability has  many implications for characterizing the sanitizer's effectiveness. In panels  b) and d)  cyan and yellow  connect to the air conditioner. The yellow dot also identifies the region above the patient's head, which forward time stays circulating at the bottom of the room. Finally, blue dots correspond to the intersection of the circulating trajectory at the foot of the bed, in the lower part of the room. Figure \ref{funM3d} presents a joint 3D composition of the slices displayed in Figures \ref{funM1} and \ref{funM2b}. 

\section{Discussions and conclusions}\label{Conclusions and discussion}
In this study, an extensive analysis was performed to examine SARS-CoV-2 air transport in a three-dimensional hospital isolation room and evaluate the effectiveness of air conditioning and sanitizers in reducing the  spread of SARS-CoV-2.
The combination of CFD simulation and LCS analysis allowed for a comprehensive understanding of the mechanisms of fluid particle transport and virus dispersion in the room. The CFD findings showed that the placement and direction of air conditioning and sanitizer devices can significantly affect the flow motion behavior in the room. LCS were computed through the method of Lagrangian Descriptors. This method  was successfully applied to find ordered transport patterns of fluid particles within the designed hospital isolation room over a specific time range of 10 seconds. LD provided a partition of the space around the patient, which allowed classifying the  type of particle evolution and extracting conclusions about the spreading of the virus. 
The results showed that LD allowed identifying the intersections of the air volume that contains the trajectories that fly from the sanitizer towards the patient's head. Similar volume intersections were identified for trajectories that fly from the sanitizer towards the patient's head and from the patient's head  towards the exhaust port. These findings demonstrated the potential of LD as a useful tool for analyzing the transport behavior of fluid particles in complex environments such as hospital isolation rooms and their ability to develop effective strategies for controlling the spread of SARS-CoV-2 and improving the safety of healthcare workers and patients.

The methodology employed in this study holds the potential for broader applications, beyond the specific context in which it has been implemented. It could effectively evaluate the efficiency of ventilation systems in interior settings and it could be extended to explore other spaces such as transport vehicles, and architectural interiors intended for collective and public use.

\section*{Acknowledgements}
The authors express their gratitude to ENDESA for providing the funding for this research, and to the Institute of Mathematical Sciences (ICMAT)  for their support and provision of time and facilities for the study.
GGS and AMM acknowledge the support of a CSIC PIE project Ref. 202250E001.  AMM, GGS and MA acknowledge the support from grant PID2021-123348OB-I00 funded by 
MCIN/ AEI /10.13039/501100011033/ and by
FEDER A way to make Europe. MA acknowledges the support from the grant CEX2019-000904-S and IJC2019-040168-I funded by: MCIN/AEI/10.13039/501100011033. Authors acknowledge participation at CSIC Interdisciplinary Thematic Platforms TELEDETECT and POLARCSIC.

\bibliographystyle{elsarticle-num}
\bibliography{SNreac}

\begin{thebibliography}{10}
\expandafter\ifx\csname url\endcsname\relax
  \def\url#1{\texttt{#1}}\fi
\expandafter\ifx\csname urlprefix\endcsname\relax\def\urlprefix{URL }\fi
\expandafter\ifx\csname href\endcsname\relax
  \def\href#1#2{#2} \def\path#1{#1}\fi

\bibitem{Setti2020}
L.~Setti, F.~Passarini, P.~De~Gennaro G.and~Barbieri, M.~Perrone, M.~Borelli,
  J.~Palmisani, A.~Di~Gilio, P.~Piscitelli, A.~Miani, Airborne transmission
  route of covid-19: Why 2 meters 6 feet of inter-personal distance could not
  be enough, International journal of Environment Research and Public health
  17~(8) (2020) 2932.
\newblock \href {https://doi.org/10.3390/ijerph17082932}
  {\path{doi:10.3390/ijerph17082932}}.

\bibitem{Verma2018}
T.~Verma, S.~Sahu, A.K.and~Sinha, Air Pollution and Control, Springer, 2018.
\newblock \href {https://doi.org/10.1007/978-981-10-7185-0\_11}
  {\path{doi:10.1007/978-981-10-7185-0\_11}}.

\bibitem{Bhatia2020}
D.~Bhatia, A.~D. Santis, A preliminary numerical investigation of airborne
  droplet dispersion in aircraft cabins, Open Journal of Fluid Dynamics 10~(03)
  (2020) 198--207.
\newblock \href {https://doi.org/10.4236/ojfd.2020.103013}
  {\path{doi:10.4236/ojfd.2020.103013}}.

\bibitem{Bhattacharya2020}
S.~Bhattacharyya, K.~Dey, R.~Paul, A.R.and~Biswas, A novel cfd analysis to
  minimize the spread of covid-19 virus in hospital isolation room, Chaos,
  Solitons and Fractals 139~(1) (2020) 110294.
\newblock \href {https://doi.org/10.1016/j.chaos.2020.110294}
  {\path{doi:10.1016/j.chaos.2020.110294}}.

\bibitem{Leonard2020}
S.~Leonard, W.~Strasser, J.~Whittle, L.~Volakis, R.~DeBellis, R.~Prichard,
  C.~Atwood, G.~Dungan, Reducing aerosol dispersion by high flow therapy in
  covid‐19: High resolution computational fluid dynamics simulations of
  particle behavior during high velocity nasal insufflation with a simple
  surgical mask, Journal of the American College of Emergency Physicians Open
  1~(4) (2020) 578--591.
\newblock \href {https://doi.org/10.1002/emp2.12158}
  {\path{doi:10.1002/emp2.12158}}.

\bibitem{Arjmandi2022}
H.~Arjmandi, M.~Amini, R.and~Kashfi, A.~Abikenari, M.A.and~Davani, Minimizing
  the covid-19 spread in hospitals through optimization of ventilation systems,
  Physics of Fluids 34~(3) (2022) 37103.
\newblock \href {https://doi.org/10.1063/5.0081291}
  {\path{doi:10.1063/5.0081291}}.

\bibitem{speetjens}
M.~F.~M. Speetjens, H.~J.~H. Clercx, H.~G.~J. F., { A numerical and
  experimental study on advection in three-dimensional Stokes flows.}, J. Fluid
  Mech. 514 (2004) 77--105.

\bibitem{wiggins2010}
S.~Wiggins, Coherent structures and chaotic advection in three dimensions, J.
  Fluid Mech. 654 (2010) 1--4.
\newblock \href {https://doi.org/10.1017/S0022112010002569}
  {\path{doi:10.1017/S0022112010002569}}.

\bibitem{niang2020}
C.~Niang, A.~M. Mancho, V.~J. Garcia-Garrido, E.~Mohino, B.~Rodriguez-Fonseca,
  J.~Curbelo, {Transport pathways across the West African Monsoon as revealed
  by Lagrangian Coherent Structures }, Scientfic Reports 10 (2020) 12543.

\bibitem{renzo2023}
R.~Bruera, J.~Curbelo, G.~Garc\'ia-S\'anchez, A.~M. Mancho, {Mixing and
  geometry in the North Atlantic Meridional Overturning Circulation },
  Geophysical Research Letters 50 (2023) e2022GL102244.

\bibitem{sahul2019}
A.~K. Arvind Kumar~Sahu1, T.~N. Verma, S.~L. Sinha1, Numerical simulation of
  air flow in multiple beds intensive care unit of hospital, International
  Journal of Automotive and Mechanical Engineering 16~(32) (2019) 6796--6807.

\bibitem{Shih1995}
W.~T.Shih, A.Shabbir, Z.Yang, J.Zhu, A new k-epsilon eddy viscosity model for
  high reynolds number turbulent flows, Computers and Fluids 24~(3) (1995)
  227--228.
\newblock \href {https://doi.org/10.1016/0045-7930(94)00032-T}
  {\path{doi:10.1016/0045-7930(94)00032-T}}.

\bibitem{COMSOL}
COMSOL Multiphysics v. 5.4, www.comsol.com. Stockholm, Sweden, 2018.

\bibitem{brez2015}
A.~Brezmes, C.~Breitkopf, Fast and reliable simulations of argon inductively
  coupled plasma using comsol, Vacuum 116 (2015) 65--72.

\bibitem{z2012}
Z.~Zhu, M.~Kaliske, An iterative method to solve the heat transfer problem
  under the non-linear boundary conditions, Heat and Mass Transfer 48~(2)
  (2012) 283--290.

\bibitem{n2017}
K.~Nadolin, I.~Zhilyaev, A reduced 3d hydrodynamic model of a shallow, long,
  and weakly curved stream, Water Resources 44~(2) (2017) 237--245.

\bibitem{m2012}
M.~Mohajerani, M.~Mehrvar, E.-M. F., Cfd analysis of two-phase turbulent flow
  in internal airlift reactors, The Canadian Journal of Chemical Engineering
  90~(2) (2012) 1611--1630.

\bibitem{Manjesh2022}
K.~Manjesh, K.~Vikash, K.~Abhinav, H.~Yadav, D.~Manas, Cfd numerical simulation
  in building drainage stacks as an infection pathway of covid-19,
  International Journal of Environmental Research and Public Health 19~(12)
  (2022) 7475.
\newblock \href {https://doi.org/10.3390/ijerph19127475}
  {\path{doi:10.3390/ijerph19127475}}.

\bibitem{Sofi2023}
S.~Anas, A.~Narjisse, E.~Abderrahman, Modelling of neutral-stratified
  atmospheric boundary layer with commercial cfd software for the horizontal
  homogeneity thermo-fluid problem, Vol. 605, Springer, Cham, 2023, p.
  649–659.
\newblock \href {https://doi.org/10.1007/978-3-031-22375-4\_52}
  {\path{doi:10.1007/978-3-031-22375-4\_52}}.

\bibitem{Cheng2022}
L.~Cheng, Y.~Yen, Cfd numerical simulation in building drainage stacks as an
  infection pathway of covid-19, International Journal of Environmental
  Research and Public Health 19~(12) (2022) 7475.
\newblock \href {https://doi.org/10.3390/ijerph19127475}
  {\path{doi:10.3390/ijerph19127475}}.

\bibitem{narjisse2021assessment}
A.~Narjisse, K.~Abdellatif, Assessment of rans turbulence closure models for
  predicting airflow in neutral abl over hilly terrain, International Review of
  Applied Sciences and Engineering 12~(3) (2021) 238--256.

\bibitem{psc2010}
Z.~Pouransari, M.~Speetjens, H.~Clercx, Formation of coherent structures by
  fluid inertia in three-dimensional laminar flows, J. Fluid Mech. 654 (2010)
  5--34.

\bibitem{iri}
I.~I. Rypina, L.~J. Pratt, P.~Wang, T.~M. \"Ozg\"okmen, I.~Mezic, Resonanace
  phenomena in a time-dependent, three-dimensional model of an idealized eddy,
  Chaos 25 (2015) 087401.

\bibitem{shadcen2005}
S.~C. Shadden, F.~Lekien, J.~E. Marsden, {Definition and properties of
  Lagrangian Coherent Structures from finite-time Lyapunov exponents in
  two-dimensional aperiodic flows}, Physica D 212 (2005) 271--304.

\bibitem{Madrid2009}
J.~A.~J. Madrid, A.~M. Mancho, {Distinguished trajectories in time dependent
  vector fields}, Chaos 19 (2009) 013111.
\newblock \href {https://doi.org/10.1063/1.3056050}
  {\path{doi:10.1063/1.3056050}}.

\bibitem{mendoza2010}
C.~Mendoza, A.~M. Mancho, Hidden geometry of ocean flows, Phys Rev Lett 105
  (2010) 038501.
\newblock \href {https://doi.org/10.1103/PhysRevLett.105.038501}
  {\path{doi:10.1103/PhysRevLett.105.038501}}.

\bibitem{mancho2013lagrangian}
A.~M. Mancho, S.~Wiggins, J.~Curbelo, C.~Mendoza, Lagrangian descriptors: A
  method for revealing phase space structures of general time dependent
  dynamical systems, Communications in Nonlinear Science and Numerical
  Simulation 18~(12) (2013) 3530--3557.
\newblock \href {https://doi.org/10.1016/j.cnsns.2013.05.002}
  {\path{doi:10.1016/j.cnsns.2013.05.002}}.

\bibitem{lopesino2017}
C.~Lopesino, F.~Balibrea-Iniesta, V.~J. Garc\'ia-Garrido, S.~Wiggins, A.~M.
  Mancho, A theoretical framework for lagrangian descriptors., International
  Journal of Bifurcation and Chaos 27 (2017) 1730001.

\bibitem{curbelo2017}
J.~Curbelo, V.~J. Garc\'ia-Garrido, C.~R. Mechoso, A.~M. Mancho, S.~Wiggins,
  C.~Niang, Insights into the three-dimensional lagrangian geometry of the
  antarctic polar vortex, Nonlin. Proc. Geophys. 24 (2017) 379--392.

\bibitem{curbelo2018b}
J.~Curbelo, C.~R. Mechoso, A.~M. Mancho, S.~Wiggins, Lagrangian study of the
  final warming in the southern stratosphere during 2002: Part ii. 3d
  structure., Climate Dynamics 53 (2019) 1277--1288.

\bibitem{curbelo2018a}
J.~Curbelo, C.~R. Mechoso, A.~M. Mancho, S.~Wiggins, Lagrangian study of the
  final warming in the southern stratosphere during 2002: Part i. the vortex
  splitting at upper levels., Climate Dynamics 53 (2019) 2779--2792.

\bibitem{craven2015lagrangian}
G.~T. Craven, R.~Hernandez, Lagrangian descriptors of thermalized transition
  states on time-varying energy surfaces, Phys Rev Lett 115~(14) (2015) 148301.
\newblock \href {https://doi.org/10.1103/PhysRevLett.115.148301}
  {\path{doi:10.1103/PhysRevLett.115.148301}}.

\bibitem{junginger2016transition}
A.~Junginger, G.~T. Craven, T.~Bartsch, F.~Revuelta, F.~Borondo, R.~Benito,
  R.~Hernandez, Transition state geometry of driven chemical reactions on
  time-dependent double-well potentials, Phys Chem Chem Phys 18~(44) (2016)
  30270--30281.

\bibitem{Agaoglou2019}
M.~Agaoglou, B.~Aguilar-Sanjuan, V.~J. Garc{\'i}a-Garrido,
  R.~Garc{\'i}a-Meseguer, F.~Gonz{\'a}lez-Montoya, M.~Katsanikas, V.~Krajňák,
  S.~Naik, S.~Wiggins, Chemical Reactions: A Journey into Phase Space, Zenodo,
  2019.
\newblock \href {https://doi.org/10.5281/zenodo.3568210}
  {\path{doi:10.5281/zenodo.3568210}}.

\end{thebibliography}
\end{document}